**Effect of an auditory static distractor on the perception of an auditory moving target**

Noa Kemp[1,2], Cynthia Tarlao[2,3], Catherine Guastavino[2,3*] and B. Suresh Krishna[1,2*]

[1] *Department of Physiology, McGill University, Montréal, QC, Canada*

[2] *Center for Interdisciplinary Research in Music Media and Technology, Montréal, QC, Canada*

[3] *School of Information Studies, McGill University, Montréal, QC, Canada*

* Joint senior authors

Correspondence: suresh.krishna@mcgill.ca

*Author-prepared version incorporating final proof corrections*

*The following article appeared in J. Acoust. Soc. Am. 160, 3003–3019 (2026) and may be found at https://doi.org/10.1121/10.0044523*

## ABSTRACT

It is known that listeners lose the ability to discriminate the direction of motion of a revolving sound (clockwise vs counterclockwise) beyond a critical velocity ["the upper limit (UL)"], likely due to degraded front–back discrimination. Little is known about how this ability is affected by simultaneously present distractor sounds, despite the real-life importance of tracking moving sounds in the presence of distractors. We hypothesized that the presence of a static distractor sound would impair the perception of moving target sounds and reduce the UL and show that this is indeed the case. A distractor on the right was as effective as a distractor at the front in reducing the UL, despite the likely importance of resolving front–back confusions. By manipulating the spectral content of both the target and distractor, we found that the UL was significantly reduced only when the distractor spectrally overlaps with the target in the frequency range relevant for front–back discrimination. Our findings form the first steps toward a better understanding of the tracking of multiple sounds in the presence of distractors.

## I. INTRODUCTION

In everyday environments, listeners are surrounded by multiple sound sources, and accurate localization is critical for navigating complex auditory scenes (e.g., avoiding an approaching vehicle). This process depends not only on identifying the location of a sound but also on tracking a moving source, or "target," in the presence of irrelevant sounds, or distractors. The mechanisms underlying static sound localization are well established; they rely on binaural cues—interaural time differences (ITDs) and interaural level differences (ILDs)—as well as monaural spectral cues shaped by head-related transfer functions (HRTFs). However, research on the perception of moving sounds has progressed more slowly than that on static localization, in part, due to the need for complex experimental setups, such as loudspeaker arrays and spatialization techniques.[1] Psychoacoustic research on the perception of moving sounds[1] has established the existence of a minimum audible movement angle, or MAMA, defined as the minimum angle a target sound needs to be moved to be distinguished from a static sound.[2] Though differences exist across studies (using different stimuli and velocities), the MAMA generally increases when velocity increases and when spectral bandwidth decreases.[1,3] Moreover, the MAMA is highly dependent on spatial position: listeners can detect much smaller angular changes at the front (0° azimuth) than on the sides (± 90° azimuth).[1] Velocity discrimination thresholds for moving sounds have also been studied, showing that Weber fractions decrease as the reference velocity increases.[4,5]

Research on auditory motion perception has mostly relied on a single sound moving at slow velocities over narrow displacements.[4] However, sound sources can travel larger distances at faster velocities; e.g., motorised vehicles or video-game audio. Studies involving larger, circular displacements have examined velocity discrimination[4] and upper limits of circular-motion perception.[6–8] For direction discrimination, the upper limit (UL) is the velocity above which clockwise (CW) and counterclockwise (CCW) motion can no longer be reliably distinguished.[6,7] Using a rotation-versus-no-rotation judgment, Féron et al.[8] reported a white-noise (WN) threshold of around 2.5 rotations per second (rot/s). Circular trajectories centered on a stationary listener allow azimuth to vary while source distance remains fixed, avoiding direct-sound Doppler shifts and the distance-related changes in attenuation and direct-to-reverberant ratio associated with approaching or receding motion. Roggerone *et al.*[6] showed that when frequencies likely needed for front–back discrimination (11.3–16 kHz range) are removed from the spectrum of the revolving sound, the UL significantly decreases. Roggerone *et al.*[6] also showed that a model based on temporal integration of a series of static sound-localization snapshots[9] could predict behavioral results very well. Although Roggerone et al. did not measure front–back confusions directly, their spectral manipulations were motivated by Langendijk and Bronkhorst,[10] who identified frequency bands contributing to front–back localization by removing direction-dependent spectral information from selected HRTF bands. Roggerone et al.[6] and the present study instead used band-stop (BS) filtering to attenuate energy in corresponding frequency bands. Thus, there is highly suggestive indirect evidence that the UL in our circular motion task is likely due to degraded front–back discrimination, attributed to a temporal integration blur of successive sound localizations at very high velocities. This is further supported by the fact that in all our previous studies on circular motion, participants reported perceiving the sound as alternating between left and right for velocities above the UL: this informs our interpretation of the relevant factor being the loss of front–back discrimination.

### A. Research questions and hypotheses

Here, we build upon these previous studies, and add a distractor to bring us one step closer to everyday auditory scenes. In experiment 1, we measured the UL for revolving targets in the presence of a static distractor (whose position was varied). We hypothesized that the presence of a WN static distractor would impair the perception of a moving target and specifically, would reduce the UL. We tested two spatial positions for the distractor—at the front, and on the right; since this is an exploratory study that uses a distractor for the first time, we chose these two locations pragmatically based on the reasoning that a distractor in front could potentially be more effective by interfering with front–back discrimination (but see Discussion, Sec. VI B). We show below that we indeed find that a static distractor can interfere with a moving target and reduce the UL, and that this effect does not appear to depend on distractor location. We hypothesized that the presence of the distractor would result in a drop in performance for UL when the distractor contained energy in the spectral region relevant for front–back location discrimination. In experiments 2–4, we manipulated the spectral content of both the target and distractor and interpret the results in terms of the relative contribution of energetic and informational masking (see Kidd *et al.*[11] for a review) to the distractor effects that we found.

**B. Prior work**

To our knowledge, no previous study has examined how a static auditory distractor affects the upper velocity limit for discriminating the direction of circular auditory motion. Previous studies have examined several other aspects of moving sound perception in the presence of static competing sounds or background noise. Using headphone simulations, Saberi et al.[12] found that static and moving distractors impaired detection of a target's motion. For larger target excursions, static distractors were less disruptive than moving distractors. Interference decreased as target–distractor frequency separation increased but persisted even across several critical bands. In auditory-only conditions using loudspeaker presentation, Marentakis and McAdams[13] found that static distractors impaired identification of some motion trajectories in a concert hall, while others showed little or no impairment; distractor effects were smaller in a studio. Gröhn[14] also studied continuous tracking of a moving virtual sound over loudspeakers in the presence of static or moving distractors. Localization was less accurate than in an earlier experiment without distractors, and listeners sometimes tracked the distractor instead of the target. Using a circular loudspeaker array, Senna et al.[15] presented sounds rotating at about 0.47–1.19 rot/s and found that reducing their signal-to-noise ratio against spatially distributed background noise made them appear slower and reduced the precision of speed judgments. Lundbeck et al. conducted three studies of movement perception. In the first,[16] they found that adding static lateral distractors impaired detection of left–right (angular) motion presented via headphones, with a greater impairment in older listeners with hearing loss than in young listeners with normal hearing. Reverberation particularly impaired near–far (radial) motion detection in the older group. In their subsequent headphone study,[17] testing whether hearing-aid signal processing could improve movement detection in older hearing-impaired listeners, simulated bilateral beamforming improved detection of both angular and radial movements in the presence of static distractors, whereas combined directional-microphone and noise-reduction processing improved radial movement detection. Their third study tested older listeners with hearing loss wearing bilateral hearing aids, using a loudspeaker array to present a simulated street scene with one moving source among static competitors.[18] Source-counting accuracy decreased as the number of simultaneous sounds increased, while movement-direction judgments depended on the target's starting position. Other studies measured the target levels needed to discriminate motion direction in noise. Using headphone simulations and reverse correlation, Park and Fine[19] related fluctuations in background-noise amplitude at different locations and times to listeners' left–right motion judgments. Their results suggested that listeners inferred motion direction by comparing sound locations at motion onset and offset. Harada et al.[20] varied interaural level differences over headphones to simulate left–right motion of a pink-noise target amid an uncorrelated pink-noise distractor. With target duration fixed at 1 s, direction discrimination required a higher signal-to-noise ratio for the smaller (50%) than the larger (100%) amplitude-change condition, with no significant difference between younger and older adults. Neither Park and Fine nor Harada et al. included a distractor-absent comparison.

Previous studies have also examined how static auditory distractors affect the perceived trajectories of moving sounds. Petropavlovskaya et al.[21] presented broadband noise distractors and targets over headphones, using interaural level differences to simulate target movement at nominal angular velocities of 1.25 rot/s or 0.625 rot/s between the midline (0° azimuth) and either side (±90°), with the target moving either toward or away from the midline. They found that the perceived start and end positions generally shifted away from the distractor, with the pattern depending on the spatial relationship between the distractor and the target trajectory. In a free-field study, Agaeva and Nikitin[22] found that a static distractor compressed the perceived horizontal trajectory of a moving target: starting points shifted in the direction of motion and endpoints shifted oppositely, with larger starting-point shifts for targets moving away from the distractor. Related free-field precedence-effect studies used a static distractor (leading signal) and a moving target (lagging signal), with the target presented as a delayed copy of the distractor. With targets traversing a 34° arc over 1 s, Agaeva[23] found that short lead–lag delays (below approximately 8–10 ms) could prevent localization of the moving target in both the horizontal and vertical planes. In another precedence-effect study, Agaeva and Nikitin[24] found that a static leading distractor compressed the perceived horizontal trajectory of a moving lagging target, with the compression weakening as the lead–lag delay increased. In a later vertical-plane study,[25] an overhead static distractor shortened the perceived trajectory mainly by shifting the starting point in the direction of motion. The trajectory expanded with increasing distractor–target onset delay but remained shorter than in quiet even at a delay of 200 ms, and the target was detected more effectively when moving away from the distractor.

Together, these findings show that competing sounds can alter motion detection, speed judgments, and perceived trajectories. Where target speeds can be expressed in angular terms, most were at or below about 60°/s; the fastest, up to about 1.25 rot/s, were in Senna et al.[15] and Petropavlovskaya et al.[21]

A handful of other studies have used a moving sound target with static distractors to examine whether target motion could induce a release from masking. In two studies, Pastore and Yost[26,27] examined whether moving a target sound presented among distractors enhances the phenomenon of spatial release from masking (SRM) that is well documented for static sounds. In SRM, a static target becomes less subject to masking by a distractor as it becomes more spatially separated from the distractor.[28] Pastore and Yost[26] investigated whether moving an auditory target among static distractors would produce a "pop out" effect such that it became better perceptible compared to when it was static. In the first study, they panned a target speech sound 40° on the azimuthal plane over loudspeakers at a velocity of 53.33°/s (using sine-law amplitude panning), in the presence of two or four masker words that were either co-located with the target or spatially separated. Participants indicated the target word, while the number and position of distractor words were manipulated, as well as whether the target sound was panned. While the target became more perceptible when spatially separated from the distractor, the movement of the target did not provide additional benefit. The same authors also reached a similar conclusion in their second study[27] measuring tone-in-noise and noise-in-noise detection thresholds in the presence of different maskers, with the target sound either static or moving at 80°/s, where they again showed that target movement alone did not lower detection thresholds. Similar results were also obtained by Davis et al.,[29] who found that small spatial separations between target and distractor speech improved recognition in both moving and static conditions, suggesting that spatial separation could explain the apparent benefit of target motion. No improvement in detection due to motion was observed in two earlier studies using targets with varying interaural time or phase differences presented in masking noise over headphones,[30,31] as well as in a free-field detection study[32] using small-amplitude azimuthal motion (at 30°/s or 90°/s) of low-/high-frequency tones and noise targets presented among two static maskers. Muñoz et al.[33] used headphone simulations with non-individualized HRTFs and found small increases in SRM with radial distractor motion in anechoic conditions, but reductions with both radial and circular distractor motion in older listeners under highly reverberant conditions.

Finally, Cho and Kidd[34] presented a speech target over headphones among two other speech distractors and showed that participants could use motion to segregate the target sound: participants could identify a target word specified only by the fact that it was moving. The word became more identifiable and reportable as its movement amplitude (2 Hz sinusoidal displacement on the azimuthal plane spanning up to 30° amplitude) increased on the azimuthal plane, reflecting the well-known improved ability to detect motion as the displacement distance increases. However, when the target location was cued before stimulus presentation, increasing the target's motion amplitude did not significantly improve speech identification.

Taken together, these studies show that competing sounds can influence auditory motion perception and that target motion can sometimes aid source segregation. Here, we examine how a static auditory distractor affects the upper velocity limit for discriminating the direction of circular auditory motion and whether this effect depends on the distractor's spatial location and spectral content.

## C. General methods

### *1. Apparatus*

All experiments were conducted in the Performance Research Laboratory (PeRL) (W7.7 m × L9.3 m × H4.95 m) at the Center for Interdisciplinary Research in Music Media and Technology (CIRMMT) in Montreal, Canada. Stimuli were presented over a circular array of 16 Genelec 8050A loudspeakers (Genelec, Iisalmi, Finland) that were evenly spaced on a circle with a 3.3 m radius; all speakers were roughly at ear level. The entire process of stimulus generation, presentation, and data collection was controlled with Max/MSP (Cycling '74, San Francisco, CA, United States). Stimuli were played on an Apple MacBook Pro (Apple, Cupertino, CA, United States) using an RME Digiface AVB sound card (RME/Audio AG, Haimhausen, Germany) connected to an RME M-32 DA Pro digital–analog converter (RME/Audio AG, Haimhausen, Germany).

We use a modification of vector base amplitude panning (VBAP) where the spatial position is computed for each sample in the azimuthal plane using two adjacent speakers.[6,35] Stimuli consisted of WN and filtered noise generated in Max/MSP with a duration no longer than 3 s; linear 10 ms fade-in and fade-out ramps were applied to prevent clicks at the onset and offset of the stimuli.

#### *2. Procedure*

In each experiment, participants were seated in the center of the loudspeaker array and asked to keep their head still and upright. On each trial, participants were asked to indicate whether the moving target sound was revolving CW or CCW by pressing the right or left arrow keys, respectively; participants could take their time before answering. Trials were grouped into blocks corresponding to each condition. In each block, the UL was measured by increasing the speed of the moving sound with correct answers and decreasing its speed with wrong answers. This was done using an adaptive two-up one-down two-alternative forced choice task with staircases designed to converge at 70.7% correct answers; we note that up and down here refer to an increase and decrease in speed, respectively, such that up makes the task harder, and down makes it easier. We used four intertwined ascending staircases with an initial speed of 1.3 rot/s for experiment 1, and 0.9 rot/s for experiments 2–4; starting speaker, rotation direction, and staircase choice were randomized for each trial. The initial step size was 15% of the initial speed, which was halved after the third and fifth reversals, based on pilot experiments and Roggerone *et al.*[6] The staircase stopped after 12 reversals or 60 trials, whichever came first. Then, the averages across the four staircases over the last four reversals were averaged to estimate the UL. For participant 17 in Experiment 1, one control staircase and two right-distractor staircases were excluded because repeated errors at low speeds caused the adaptive procedure to request a negative speed.

#### *3. Data analysis*

To statistically analyze the ULs in each condition, we conducted a one-way repeated-measures analysis of variance (ANOVA) with *condition* as the within-subjects factor, treating *participant* as the repeated measure. A Greenhouse–Geisser correction was applied if Mauchly's test showed a statistically significant deviation from sphericity, and *post hoc* comparisons were conducted when appropriate. All statistical tests were performed with JASP 0.19.1 (JASP, Amsterdam, The Netherlands);[36] other analyses and plots were made with MATLAB 2026a (The MathWorks Inc., Natick, MA, United States).

## II. EXPERIMENT 1: DOES A STATIC DISTRACTOR AFFECT THE UL FOR A MOVING TARGET?

### A. Methods

#### *1. Participants*

We recruited 18 participants (11 female, 6 male, 1 non-binary) between 20 and 73 years of age; mean ($M$) = 30.4, median = 27.5, inter-quartile range (IQR) = 9, first quartile (Q1) = 22.25, third quartile (Q3) = 31.25. They were musically trained or worked in the field of sound, and reported normal hearing. They were compensated $15/h for their time.

#### *2. Procedure*

Experiment 1 lasted 1 h in a single session and was divided into four blocks. The first block was a practice block that lasted up to 3 min and included feedback; the stimulus in this block consisted of the moving target stimulus with all the possible conditions in each direction. The following three blocks lasted up to 15 min each. Each block had one of three conditions: a distractor at the front, a distractor on the right, or no distractor (see Fig. 1), with their order counterbalanced across participants.

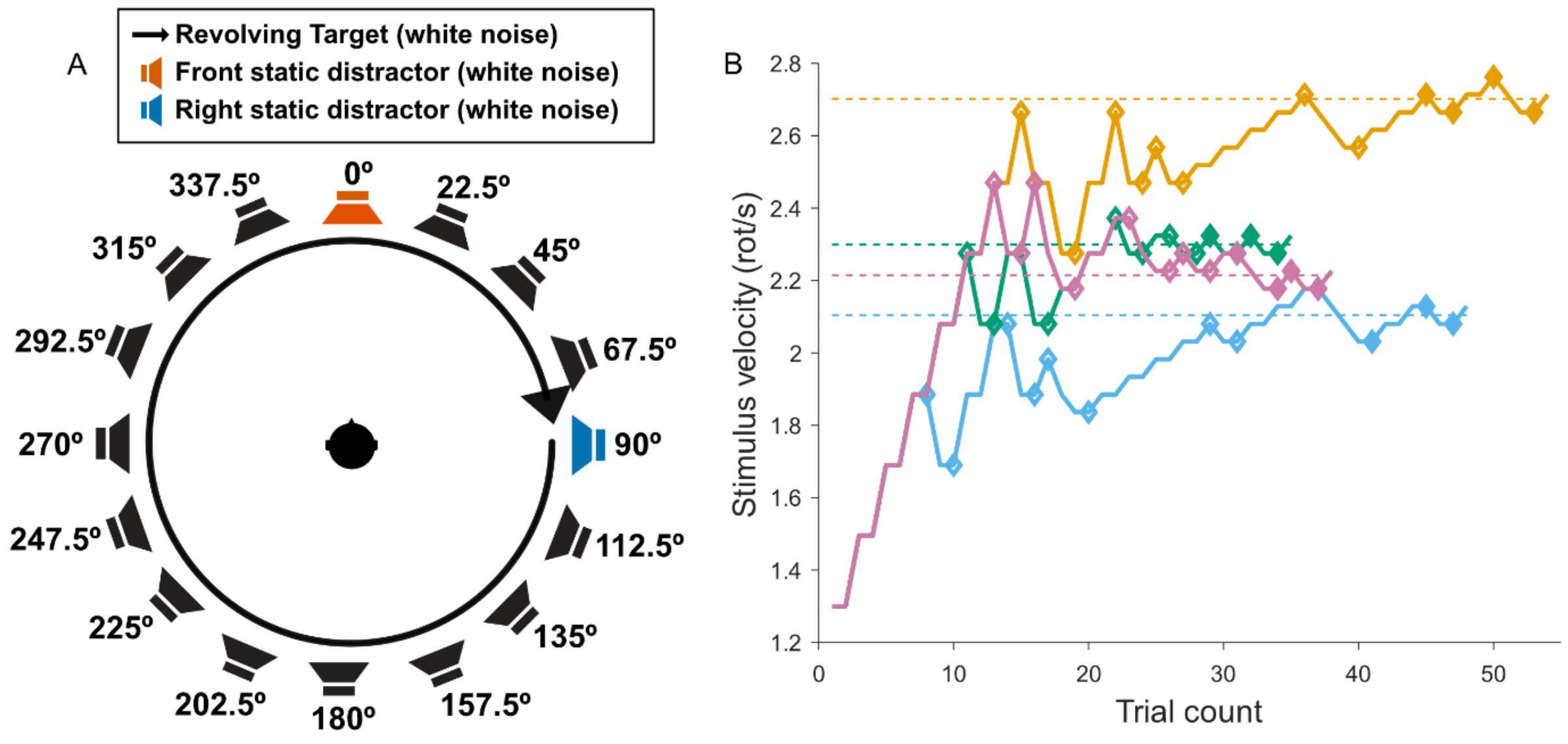

FIG. 1. Experimental setup of experiment 1 and an example staircase. (a) Loudspeaker setup: participant head at the loudspeaker array center and equidistant (3.3 m) from each of the 16 loudspeakers in the array. The distractor was presented either in front at 0° or on the right at 90°, while the target revolved around the listener using VBAP. (b) Example of the two-up one-down adaptive method used to measure the UL from one participant. The four interleaved staircases used are shown in different colors. The *y* axis shows the stimulus velocity used for each trial. The diamond signs indicate reversals, and the average of the last four reversals (filled diamonds) was used as the threshold for that staircase (indicated by the dashed line).

### *3. Stimuli*

We tested three conditions involving a revolving target and a static distractor:

• Control condition: measuring the UL of a revolving target alone.

• Distractor at the front: measuring the UL of a revolving target in the presence of a static distractor at the front.

• Distractor on the right: measuring the UL of a revolving target in the presence of a static distractor at the right.

We chose this design so that comparing the UL in the distractor conditions with that in the control would indicate whether the distractor has an effect on task performance. Similarly, comparing the UL in the two distractor conditions would indicate whether the spatial position of the distractor had an effect on task performance.

The sound pressure level (SPL) of each distractor and target was measured with a B&K 2250 sound level meter (Brüel & Kjaer, Nærum, Denmark) positioned at the center of the room, where the participant would be. For the revolving target, it was played continuously until the A-weighted decibels value stabilized.

Target: The target was a 42.5 dBA WN, generated using the WN object in Max/MSP at a sampling rate of 48 kHz (from the object "noise"). The initial speed of the target stimuli was chosen to be 1.3 rot/s (in line with Roggerone *et al.*[6]) The stimulus duration was computed on each trial via an implementation of a two-step process: first, a random duration between 2 and 3 s was chosen. Second, this number was compared to the duration required for the stimulus on that trial to complete 2, 2.5, or 3 revolutions (with the number of revolutions also chosen randomly), and then the minimum of those two durations was chosen for that trial. This two-step process enabled us to ensure that the stimulus duration was within a similar range on each trial (with a maximum of 3 s) and faster stimuli revolved 2, 2.5, or 3 times during the presentation. In this design, on average, the stimulus revolving at greater velocity will be presented for a shorter duration, since it will take less time to complete 2, 2.5, or 3 revolutions. We also note that if the speed is lower than 0.5 rot/s, the target may not perform a full revolution on some trials, and therefore the UL estimation is not accurate below this threshold.

Distractor: The distractor was also WN but uncorrelated with the target WN. The position of the distractor was either at the front, or on the right of the participant (see Fig. 1). The gain of the distractor was the same across spatial position, and the measured level was slightly higher (40.5 dBA) when presented in front compared to the right (40 dBA). The distractor had the same duration, onset, and offset time as the target.

**B. Results**

Each participant listened to a revolving target, either by itself (control condition), or in the presence of a distractor (either in front or on the right), and indicated the perceived direction of the target. For each participant (except for participant 17 in whom we excluded three invalid staircases, see Sec. I C), we averaged over the last four reversals of the four staircases to estimate their UL per condition (see Fig. 1 for staircase example and the setup), and averaged the UL across participants per condition (Fig. 2).

We conducted a one-way repeated-measures ANOVA with *participant* as the repeated measure to evaluate whether the UL differed across conditions. The assumption of sphericity was not rejected [Mauchly's test $\chi^2(2) = 1.96$, $p = 0.375$]. We found significant differences between the ULs for each condition: $F(2,34) = 31.94$, $p < 0.001$. The *condition* factor explained $\omega^2 = 22.8\%$ of the variance in UL. *Post hoc* comparisons using the Bonferroni correction show a significant difference between the front ($M = 1.73$ rot/s) and control conditions [$M = 2.24$ rot/s, mean difference = 0.51, 95% confidence interval (CI) (0.36, 0.67), $t(17) = 8.66$, $p < 0.001$, Cohen's $d = 1.20$] and between right ($M = 1.75$ rot/s) and control conditions [mean difference = 0.49, 95% CI (0.29, 0.70), $t(17) = 6.39$, $p < 0.001$, Cohen's $d = 1.15$], showing that the presence of the distractor significantly reduces the UL, regardless of its spatial position. However, no significant difference between the front and right conditions was found [mean difference = –0.02, 95% CI (–0.23, 0.19), $t(17) = -0.26$, $p = 1.000$, Cohen's $d = -0.049$], showing that the performance drop is not significantly affected by the spatial position of the distractor.

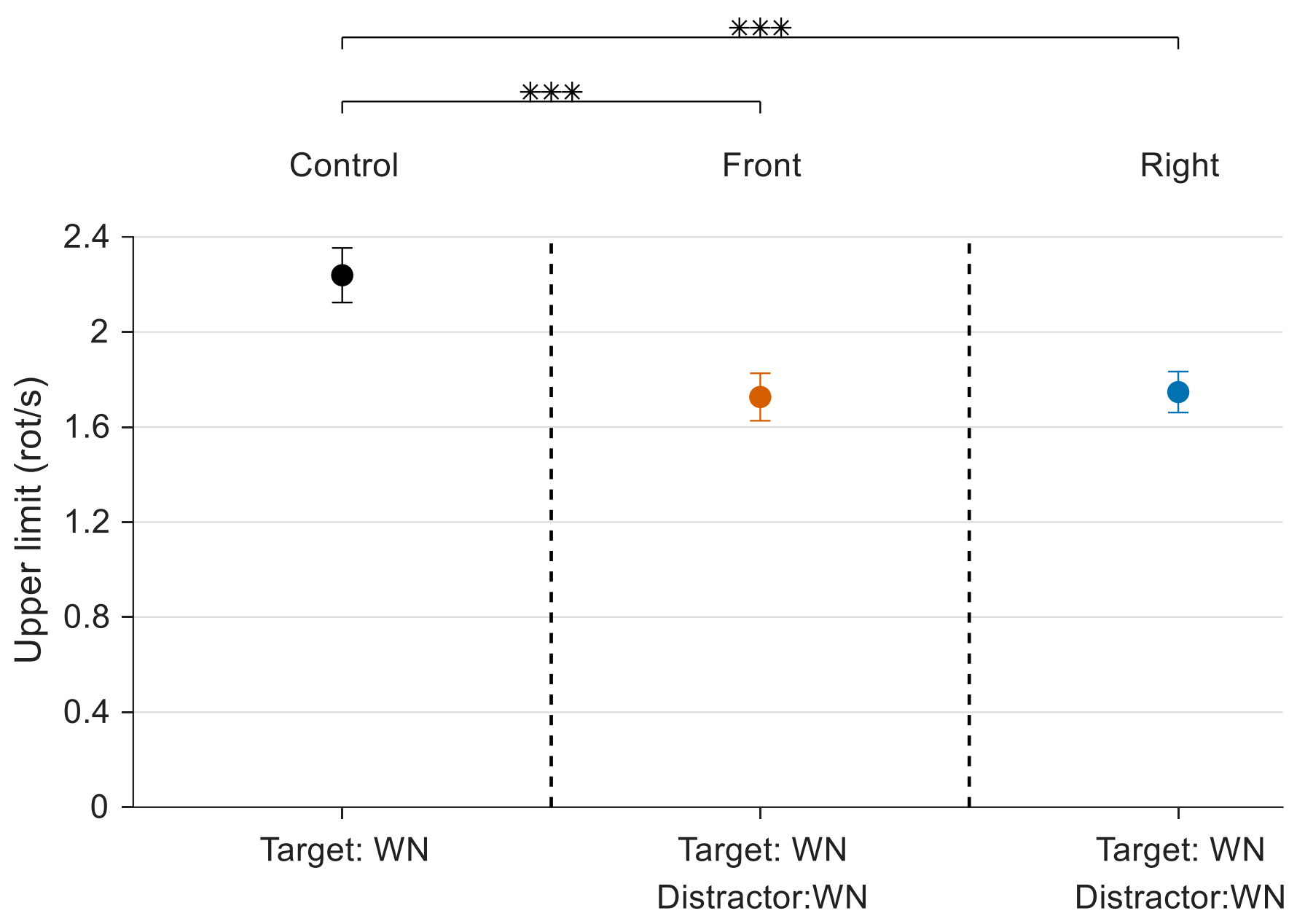

FIG. 2. Comparing UL means across participants for each condition in experiment 1: the presence of a distractor reduces the UL. Error bars are SEM. Black symbol represents the control condition with the revolving target only, with no distractors; red symbol represents the distractor presented at 0° (in front); blue symbol represents the distractor presented at 90° (on the right). Asterisk key: * Bonferroni-adjusted $p < .05$; ** $p < .01$; *** $p < .001$.

In each participant and in each plot, the within-subject variation (indicated by the standard errors of the mean (SEM) across the staircases) is relatively small compared to the across-condition effect, even though our experiment was not designed to show effects in individual participants (Fig. 8 in the Appendix). We also did not detect any dependence of per-participant effect-size on age; none of the correlations of effect-size with age were statistically significant (Fig. 8 in the Appendix).

The observed distracting effect in experiment 1 does not allow us to evaluate the spectral regions that are important for distractor interference. Accordingly, we designed experiment 2 to address this issue.

## III. EXPERIMENT 2: VARYING THE SPECTRAL OVERLAP

In experiment 1, we showed that the UL for a revolving target in the presence of a distractor significantly decreases, regardless of the spatial position of the distractor. In experiment 2, we manipulated the spectral overlap between the target and the distractor. Specifically, it has been shown that for revolving target sounds with no distractor present, removing frequency content in the 11.3–16 kHz range, which is most relevant for front–back discrimination, reduces the UL.[6] We hypothesized that the distractor would affect the UL if it contained energy in this relevant frequency range. Further, if a distractor designed to be specifically without content in this frequency range failed to reduce the UL, then there would be a lack of evidence for pure informational masking in that condition. In experiment 2, we therefore varied the overlap between the target and distractor spectral content, focusing on the contribution of the relevant frequency region for front–back discrimination (i.e., 11.3–16 kHz). Specifically, we used a broadband noise target with a spectral gap ("BS") between 5.7 and 8 kHz (for details, see Methods and Table I) by itself as the control condition to define the UL without distractors. We then used three distractor conditions, where the distractor was a static bandpass (BP) stimulus presented at the front of the listener, and whose frequency content varied between the three conditions. Specifically, the distractor's spectral content either (a) overlapped with the target in a frequency band that was not relevant for front–back discrimination (3.1–5.1 kHz): the overlap in non-relevant region (overlap-in-NRR) condition, or (b) fell within the spectral gap of the target (5.7–8 kHz) and thus had minimal overlap: the minimum-overlap condition, or (c) overlapped with the target in the relevant frequency region for front–back discrimination (i.e., 11.3–16 kHz): the overlap in relevant region (overlap-in-RR) condition. A description of the hypotheses tested by each condition can be found in Table I.

TABLE I. Summary of conditions and hypotheses. Overlap-in-NRR: There is an overlap in spectra between the target and the distractor introducing possible energetic masking. The frequency region affected should not interfere with front–back discrimination processes (Roggerone *et al.*[6]). Minimum-overlap: There is minimal spectral overlap between the target and the distractor so there should be minimal-to-no energetic masking in this condition. The frequency bands targeted should not interfere with front–back discrimination.[6] Overlap-in-RR: Spectra of the target and distractor overlap, introducing the possibility of energetic masking. The spectral overlap occurs in the 11.3–16 kHz band, shown to be relevant for front–back discrimination and likely for determining the UL.[6] The presence of the distractor in all presented conditions (except the control) may itself introduce informational masking. Interference in the minimum-overlap condition would reveal an effect of informational masking. Interference in the overlap-in-NRR condition would reflect informational masking and / or energetic masking (from a non-relevant region). Interference in the overlap-in-RR condition would reflect informational masking and / or energetic masking (from a relevant region).

| **TARGET** | **DISTRACTOR** | **Hypothesized effect of distractor** | | | **Expected performance** |
|---|---|---|---|---|---|
| | | **Informational masking** | **Energetic masking** | **Affects front/back discrimination** | |
| Gap in NRR (BS 5.7 – 8 kHz) | Control (No distractor) | | | | Comparable to WN (Roggerone 2019) |
| | Overlap in NRR (BP 3.1 – 5.1 kHz) | Yes | Yes, but in a non-relevant region | No | Comparable to control |
| | Minimum overlap (BP 5.7 – 8 kHz) | Yes | No | No | Comparable to control |
| | Overlap in RR (BP 11.3 – 16 kHz) | Yes | Yes, and in a relevant region | Yes | Significantly lower than control |

## A. Methods

### *1. Participants*

In experiment 2, we recruited a new set of 18 participants (10 women, 7 men, 1 non-binary), between 18 and 54 years of age; $M = 33$, median = 30.5, IQR = 17, Q1 = 25.25, Q3 = 42.25.

### *2. Procedure*

Experiment 2 lasted 1 h and 15 min and was divided into five blocks. The first block was a 3 min practice block with feedback, with the target revolving at 0.9 rot/s and each condition presented twice (CW and CCW). The following four blocks, corresponding to each of the four conditions, lasted up to 15 min each (Table I). We counterbalanced the order of presentation of conditions across participants.

### *3. Stimuli*

As described above, Roggerone *et al.*[6] showed that the 11.3–16 kHz frequency region is crucial for the UL task: when this frequency region is removed from the target spectrum, the UL significantly decreases (i.e., revolution direction discrimination is impaired at lower velocities). We designed a broadband revolving target with a spectral gap obtained by filtering out specific frequencies in a WN with an eighth-order Butterworth BS filter using the blocks "filterdesign" and "cascade", as in Roggerone *et al.*[6] The target remained the same across all four conditions. We also designed three different distractors (used in three different conditions) in order to gain insight into the spectral regions relevant for distractor interference. The spectral bandstop of the target was from 5.7–8 kHz, a frequency range that has been shown to not affect the UL in Roggerone *et al.*[6]

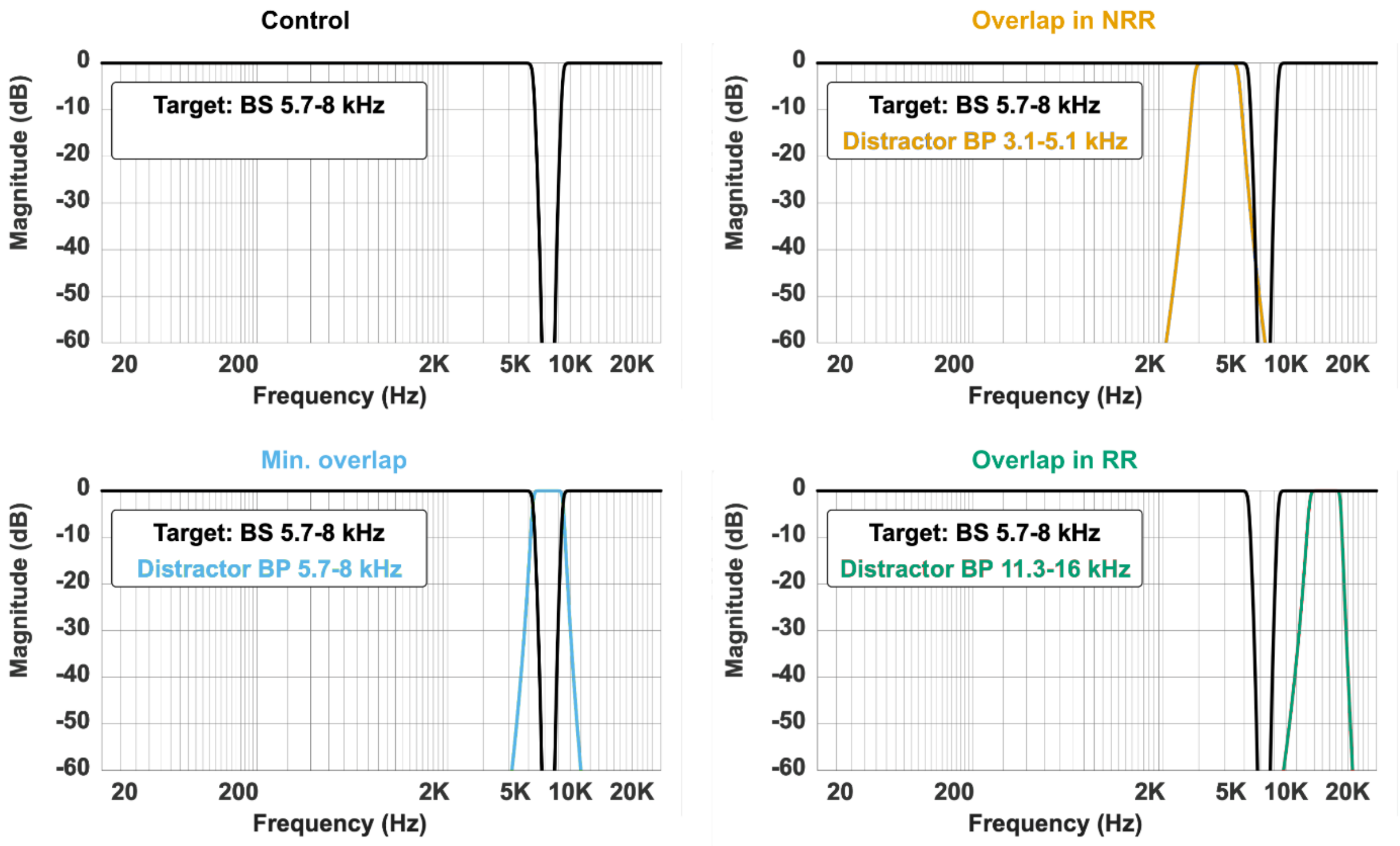

FIG. 3. Filters applied to WN for each condition in experiments 2–4. Target filter is in black; distractor filters are in colors. Orange (overlap-in-NRR): distractor overlaps in a non-relevant frequency region. Blue (minimum-overlap): distractor minimally overlaps with spectral content of the target. Green (overlap-in-RR): distractor overlaps in a relevant frequency region.

Each distractor was a static BP noise generated from WN and filtered using eighth-order Butterworth filters (Fig. 3). The BP noises contained energy in a different range in each of the three conditions: 3.1–5.1, 5.7–8, and 11.3–16 kHz (see Fig. 3 for filter responses). To keep the number of experimental trials manageable, and since there was no effect of the distractor spatial position in experiment 1, the distractor for each condition was presented from the loudspeaker at the front. The stimulus parameters (like duration) that are not explicitly specified here for experiment 2 were the same as for experiment 1.

Since the BP distractors spanned different frequency ranges, we set out to match their loudness (in sones) according to ISO 532-1[37] (Zwicker loudness model) with the function acousticLoudness from the Audio Toolbox in MATLAB (R2024B).[38] Calibrations were initially made to equate the Zwicker loudness of the BP distractors with that of the WN distractor from experiment 1 as reference. However, this calibration based on the Zwicker model produced unexpectedly high loudness values for the BP distractors, leading to high SPL, to an extent that made it difficult to conduct the experiment. To address this, we conducted a loudness matching whereby three experienced participants from our laboratory (who did not take part in any of the experiments) adjusted the loudness of each of the distractors until they were just perceptible in the presence of the target revolving at low velocity. We applied the average gain across the three participants (shown in Fig. 7 with SPLs used shown in Table II in the Appendix) in experiment 2 to ensure that the distractor was just audible in each condition.

### B. Results

We computed the UL of each participant for each condition and plotted the mean UL across participants for each condition (Fig. 4). A one-way repeated-measures ANOVA was conducted to evaluate whether there were differences in ULs depending on the condition. The assumption of sphericity was not rejected [Mauchly's test $\chi^2(5) = 6.28$, $p = 0.281$]. The ANOVA revealed a significant difference between ULs across conditions [$F(3,51) = 3.30$, $p = 0.028$]. The condition factor explained $\omega^2 = 0.7\%$ of the variance in UL. *Post hoc* comparisons using the Bonferroni correction showed a significant drop in performance for the overlap-in-RR condition ($M = 1.57$ rot/s) compared to the control condition [$M = 1.71$ rot/s, mean difference = 0.13, 95% CI (0.004, 0.27), $t(17) = 3.08$, $p = 0.041$, Cohen's $d = 0.28$]. Other comparisons showed no significant differences with the control condition: there was no significant difference for the overlap-in-NRR condition ($M = 1.65$ rot/s) compared to the control condition [mean difference = 0.06, 95% CI (–0.06, 0.17), $t(17) = 1.48$, $p = 0.944$, Cohen's $d = 0.12$] or for the minimum-overlap condition ($M = 1.66$ rot/s) compared to the control condition [mean difference = 0.04, 95% CI (–0.07, 0.15), $t(17) = 1.19$, $p = 1.000$, Cohen's $d = 0.09$]. The comparisons of the non-control conditions with each other also did not show any significant difference (all $p$-values > 0.165; see Fig. 4). Finally, as was the case for experiment 1, we did not find any statistically significant correlations of effect-size with age (Fig. 9 in the Appendix).

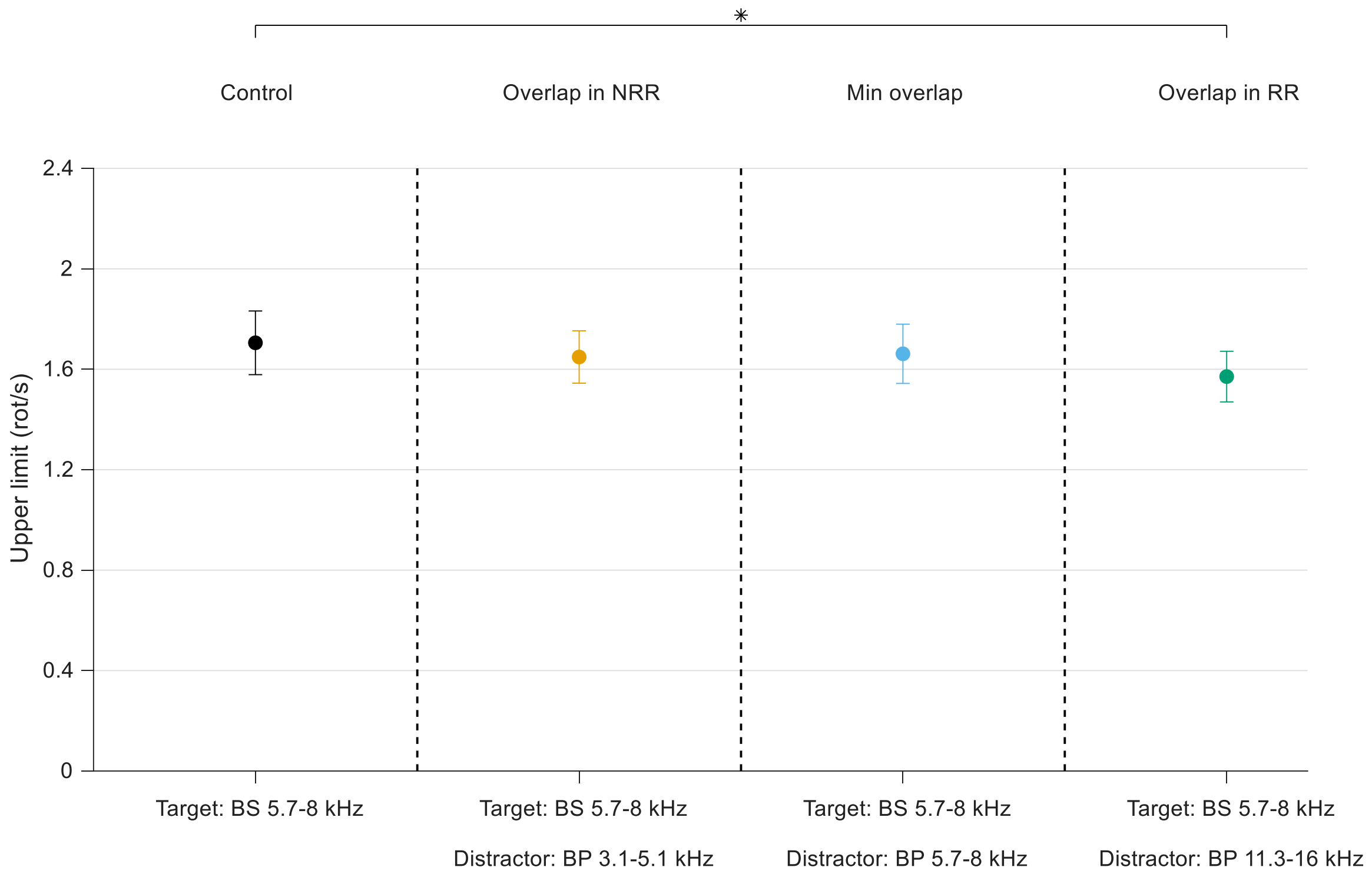

FIG. 4. Comparing UL means across participants for each condition in experiment 2: only a distractor with overlap in the relevant region reduced the target UL. Error bars are SEM. Control condition is with the revolving target only (no distractor). Orange symbol represents target and distractor overlap in spectra in a non-relevant region. Blue symbols represent minimal overlap in target and distractor spectra. Green symbols represent overlap in spectra in a relevant region for the UL.

Thus, when using BP distractors at a just-perceptible loudness level, the UL was only affected when the distractor contained energy in the frequency band (11.3–16 kHz) that had previously been shown to be relevant for this task and for front–back discrimination. We did not find any significant differences when comparing the other conditions with the control condition. This suggests that neither energetic masking in a non-relevant region, nor informational masking in the condition with minimum-overlap, leads to an impairment of the UL on our task.

We adjusted the loudness of the three distractors in this experiment to be just perceptible. With this design, the effect on the UL of the static distractor with overlap in the relevant region was statistically significant, but quite small. We therefore proceeded to conduct an additional experiment where, rather than adjusting the distractor levels in the three conditions to be just perceptible, we adjusted them to be matched in loudness (using the Zwicker loudness model) to the distractor with overlap in the non-relevant region in experiment 2.

## IV. EXPERIMENT 3: INCREASING THE LEVELS OF THE DISTRACTORS

This experiment is the same as experiment 2, except that we adjusted the distractor level in all three conditions to be matched in loudness (using the Zwicker loudness model) to the Zwicker loudness of the distractor with overlap in the non-relevant region in experiment 2. Relative to Experiment 2, this increased the measured A-weighted SPL of the distractor by 3.8 dB in the minimum-overlap condition and by 10.8 dB in the overlap-in-RR condition (see Tables II and III in the Appendix).

## A. Methods

### *1. Participants*

In experiment 3, we recruited 18 participants; however, one participant did not complete the experiment because they could not keep their head centered during testing; the 17 remaining participants (7 women, 9 men, 1 non-binary) were 20 to 58 years of age (M = 34.4, median = 28, IQR = 25, Q1 = 23, Q3 = 48). Two of these participants also participated in experiment 2.

### *2. Procedure*

The procedure was the same as in experiment 2.

### *3. Stimuli*

Loudness matching (to the loudness of the 3.1–5.1 kHz distractor used in experiment 2) was computed for the distractors with energy in the 5.7–8 and 11.3–16 kHz bands using the Zwicker Loudness model (according to ISO 532-1) using the acousticLoudness function from the Audio Toolbox in MATLAB (R2024b).[38] The 3.1–5.1 kHz distractor lies in a higher auditory sensitivity range compared to the other two distractors, leading to the other distractors having higher SPLs after matching. The SPL values measured in the room are reported in Table III in the Appendix. The stimulus parameters (like duration) that are not explicitly specified here for experiment 3 were the same as for experiment 2.

## B. Results

We computed the UL of each participant for each condition and plotted the mean UL across participants for each condition (see Fig. 5). A one-way repeated-measures ANOVA was conducted to evaluate whether there were differences in ULs depending on the condition. The assumption of sphericity was rejected [Mauchly's test $\chi^2(5) = 20.29$, $p = 0.001$], so the results are reported with the Greenhouse–Geisser correction ($\varepsilon = 0.573$). There were significant differences between the ULs across conditions [$F(1.718, 27.482) = 16.32$, $p < 0.001$]. The *condition* factor explained $\omega^2 = 17.2\%$ of the variance in UL. *Post hoc* comparisons using the Bonferroni correction show that performance significantly drops in the overlap-in-RR condition (where there is a spectral overlap between the target and the distractor in a region relevant to front–back discrimination) with a mean UL of 1.40 rot/s, when compared to all three other conditions: control condition [1.84 rot/s, mean difference = 0.44, 95% CI (0.18, 0.70), $t(16) = 5.03$, $p < 0.001$, Cohen's $d = 1.26$], minimum-overlap condition [1.72 rot/s, mean difference = 0.32, 95% CI (0.07, 0.57), $t(16) = 3.79$, $p = 0.010$, Cohen's $d = 0.93$] and overlap-in-NRR condition [1.72 rot/s, mean difference = 0.32, 95% CI (0.09, 0.55), $t(16) = 4.23$, $p = 0.004$, Cohen's $d = 0.92$]. None of the other comparisons yielded a statistically significant difference. The UL was not significantly different between the control condition and both the overlap-in-NRR condition [mean difference = 0.12, 95% CI ( –0.04, 0.28), $t(16) = 2.24$, $p = 0.238$, Cohen's $d = 0.341$] and the minimum-overlap condition [mean difference = 0.12, 95% CI (0, 0.24), $t(16) = 2.93$, $p = 0.059$,

Cohen's $d = 0.34$]. Finally, as was the case for experiments 1 and 2, we did not find any statistically significant correlations of effect-size with age (Fig. 10 in the Appendix).

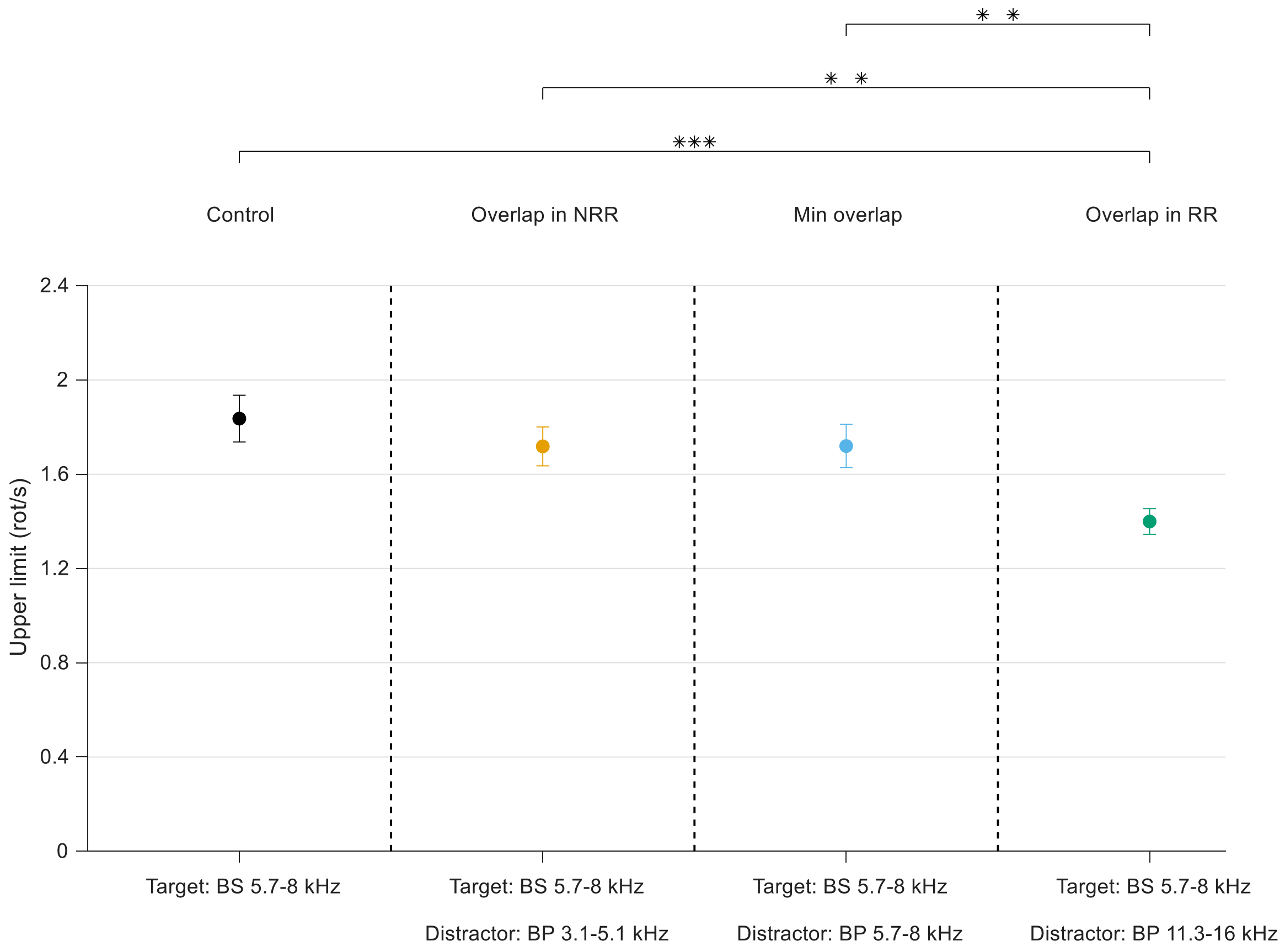

FIG. 5. Comparing UL means across participants for each condition in experiment 3: only the distractor with overlap in the relevant region reduced the target UL. Figure format as in Fig. 4.

The results of experiment 3 confirm those from experiment 2: only the distractor with overlap-in-RR reduced the UL. As in experiment 2, the UL was only affected when the distractor contained energy in the frequency band (11.3–16 kHz) that had previously been shown to be relevant for this task.

## V. EXPERIMENT 4: DISTRACTOR ON THE RIGHT

Finally, since we had only conducted experiment 3 with the distractor at the front, we replicated it with the distractor on the right in experiment 4.

### A. Methods

The procedure and stimuli are the same as in experiment 3; however, the distractor is now positioned at the right of the participant, in the same fashion as in experiment 1. The levels of the distractor are equivalent to experiment 3. This is because the SPL of the overlap-in-NRR distractor reference stimulus was similar for a distractor at the front and on the right (see Fig. 7 in the Appendix).

In this experiment, we recruited 18 participants; however, the data from one participant had to be dropped due to exceptionally low performance (such that the staircases rapidly converged to zero velocity). The 17 remaining participants (5 women, 11 men, and 1 preferred not to say) were 21–58 years of age ($M$ = 30.6, median = 25, IQR = 13, Q1 = 23, Q3 = 36), and one of them had participated in experiments 2 and 3.

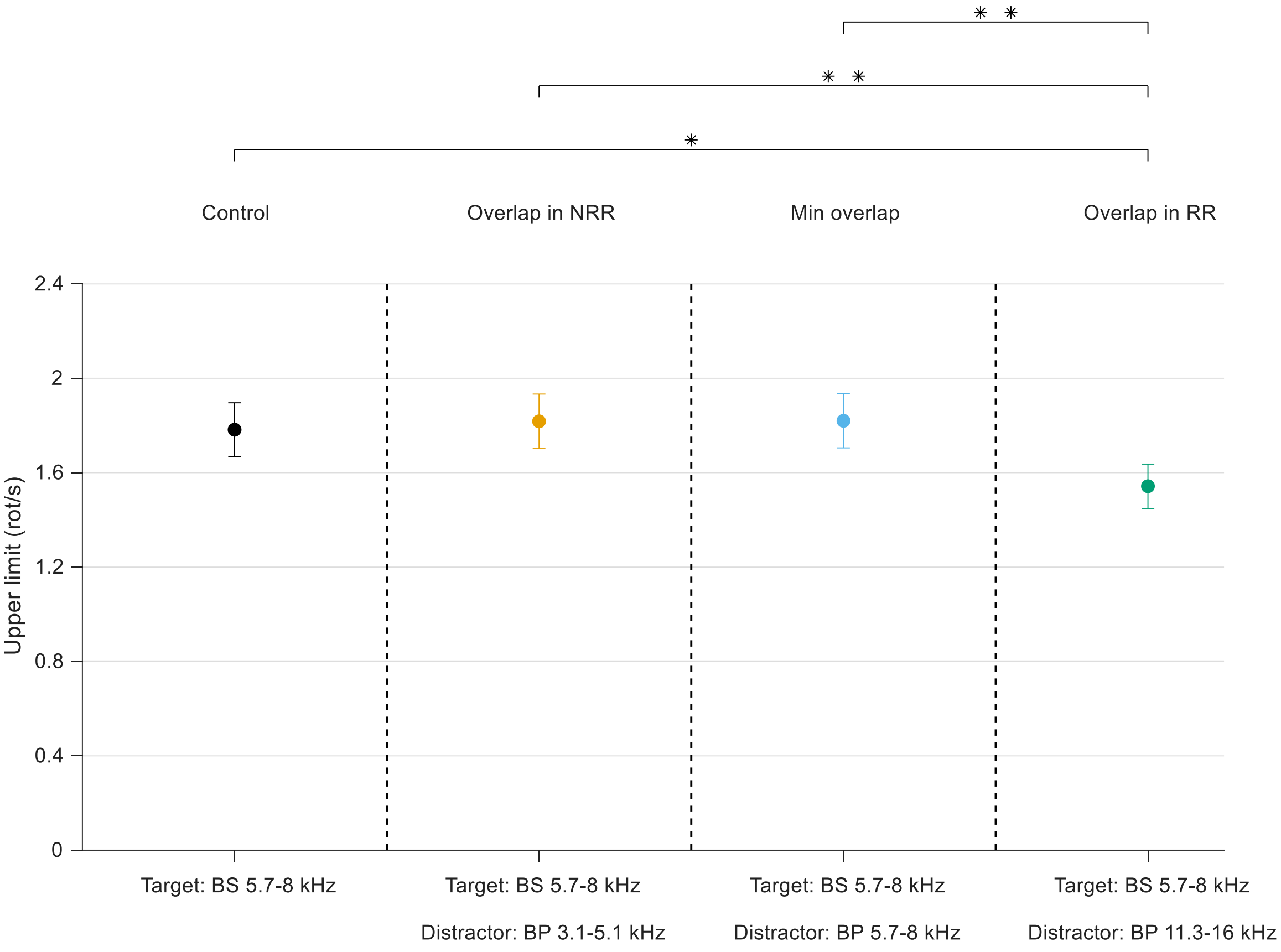

FIG. 6. Comparing UL means across participants for each condition in experiment 4 (the distractor is on the right); only the distractor with overlap in the relevant region reduced the target UL. Figure format as in Fig. 4.

## B. Results

The results from experiment 4 with the static distractor to the right of the participant were very similar to those from experiment 3 with the static distractor in front of the participant. As for the previous experiments, we computed the UL of each participant for each condition (Fig. 6) and a one-way repeated measures ANOVA was conducted to evaluate whether there were differences in ULs across conditions. The assumption of sphericity was not rejected [Mauchly's test $\chi^2(5) = 4.29$, $p = 0.509$]. There were significant differences between the ULs across conditions [$F(3,48) = 7.92$, $p < 0.001$]. The *condition* factor explained $\omega^2 = 5.3\%$ of the variance in UL. *Post hoc* comparisons using the Bonferroni correction show that performance significantly drops in the overlap-in-RR condition with a mean UL of 1.54 rot/s, when compared to all three other conditions: control condition [M = 1.78 rot/s, mean difference = 0.24, 95% CI (0.01, 0.47), $t(16) = 3.13$, $p = 0.038$, Cohen's $d = 0.53$], minimum-overlap condition [$M = 1.82$ rot/s, mean difference = 0.28, 95% CI (0.07, 0.49), $t(16) = 3.97$, $p = 0.007$, Cohen's $d = 0.61$] and overlap-in-NRR condition [$M = 1.82$ rot/s, mean difference = 0.27, 95% CI (0.07, 0.48), $t(16) = 3.95$, $p = 0.007$, Cohen's $d = 0.61$]. None of the other comparisons yielded a statistically significant difference. The UL was not significantly different between the control condition and both the overlap-in-NRR condition [mean difference = –0.04, 95% CI (–0.19, 0.12), $t(16) = -0.67$, $p = 1.000$, Cohen's $d = -0.08$] and the minimum-overlap condition [mean difference = –0.04, 95% CI (–0.22, 0.14), $t(16) = -0.64$, $p = 1.000$, Cohen's $d = -0.08$]. Finally, as was the case for experiments 1–3, we did not find any statistically significant correlations of effect-size with age (Fig. 11 in the Appendix).

## VI. DISCUSSION

Few studies have addressed the ability to track moving sound sources in the presence of distractors. Here, we explore how static distractors influence the discrimination of motion direction (CW or CCW) of a revolving target. We focus on the UL, which is the velocity threshold above which listeners lose the ability to track the direction of circular motion. This threshold is likely related to front–back confusion, and it has been shown that the UL is affected most by frequencies that contribute to front–back location discrimination.[6] We hypothesized that the presence of a static distractor would impair the perception of a moving target and would reduce the UL. Specifically, we hypothesized that the distractor would interfere with target motion perception if it contained energy in the spectral region that has previously been shown to be relevant for target front–back location discrimination. In experiment 1, we indeed found that the presence of the distractor, regardless of its spatial position (in the front or on the side), impaired performance and reduced the UL.

### A. Distractor effect depends on distractor energy at task-relevant frequencies

Roggerone *et al.*[6] showed that the auditory limitations leading to the UL likely stem from the inability to disambiguate whether the sound is at the front or the back above a certain velocity; this limitation is referred to as front–back confusion. Above this velocity, the ability to resolve front–back confusions is limited by the temporal integration of successive localization snapshots (no shorter than ~ 300 ms), resulting in a decrease in front–back discrimination. Front–back discrimination relies on monaural spectral cues, including the head-related transfer function, and Roggerone *et al.*[6] showed that the frequency region relevant to front–back discrimination, in the 11.3–16 kHz range, is crucial for rotation direction discrimination: when these frequencies are removed from the spectra, the UL significantly decreases.[6] In experiments 2–4, we aimed to better identify the spectral regions relevant for distractor interference. To do this, we manipulated the spectral content of both the target and the distractors, and used two different ways to match the loudness of the three distractors; we also varied the spatial location of the distractor (front vs right). In all these cases in experiments 2–4, regardless of the spatial location of the distractor, we found that distractor presence significantly reduces the UL only when the distractor spectrally overlaps with the target in the frequency range relevant for front–back discrimination, when compared to the control (experiment 2) and when compared to all three other conditions (experiments 3 and 4). This finding is thus consistent with the findings in Roggerone *et al.*,[6] showing that the frequency region from 11.3 to 16 kHz is crucial for the task. Additionally, the spectral overlap in the 3.1–5.1 kHz region did not affect performance, which is also in line with the findings in Roggerone *et al.*[6] The absence of a distracting effect in this condition, as well as in the minimum-overlap condition, also indicates that there was no evidence for informational masking by the distractor in these conditions. Furthermore, we found that adjusting the levels of the distractor to match them in Zwicker loudness, also only affected performance when there is an overlap in the relevant frequency region, regardless of the spatial position of the distractor. This confirms that performance highly depends on the distractor containing spectral energy in the relevant frequency region needed for front–back discrimination. The distractor with spectral energy in the relevant region could exert its effect either via energetic masking of target information, or potentially via informational masking that is higher in this overlap-in-RR condition, because the participants pay attention to the relevant frequency region to perform the task and therefore integrate distractor information into their perceptual decision-making mechanisms when the distractor has energy in this relevant frequency region. Differentiating between these mechanisms is non-trivial,[11,39] and may require new experiments that attempt to carefully manipulate auditory attentional allocation or a much better understanding of the energetic interactions relevant to this task.

**B. Similar distractor effects from the front and at the right**

We used two spatial locations for the static distractor; front (0° in azimuth) in experiments 1–3, and right (90° azimuth) in experiments 1 and 4. We observed no effect of the spatial position of the distractor in experiment 1 and when examining experiments 3 and 4. Previous research on SRM (mainly using static targets), established that a release from masking occurs when a target spatially separates from a distractor. Our experiments show both that (a) only distractors with energy in the relevant frequency region for front–back discrimination impair target perception and (b) this effect is similar for distractors at the front and on the right. One could argue that this is puzzling, since SRM indicates that the distractor should mask the revolving target only for a certain spatial width (or angle) around the distractor location, and if it is front–back discrimination that is key for the discrimination of rotation direction, a distractor at the front would be expected to cause a greater reduction of the UL. We propose two possible explanations: (a) at the high velocities around the UL, SRM may not be present, consistent with the fact that performance in spatial tasks decreases at high velocities (the MAMA increases with velocity[1] and direction discrimination worsens as velocity increases[6,7]), which could suggest that SRM would also worsen at high velocities, but previous studies have only investigated SRM with slow velocities[26,27] and (b) alternatively, SRM could indeed be present and the distractor at the front could produce a greater masking effect, but this effect could be compensated by an additional energy-based cue when the distractor is at the front: greater summation between target and distractor could lead to a perceptibly louder stimulus when the target is near the distractor, thus providing an additional cue for front–back discrimination. Additionally, it is also still possible, though we consider it unlikely, that despite the strong indirect suggestive evidence that front–back discrimination is a key factor determining the UL, it may not be operating in this task in the manner that we have envisaged. Further work will be needed to disambiguate between these and other explanations.

## C. Study limitations

The first methodological limitation of this study is the lack of audiometric testing. We relied on the participants' self-reported normal hearing at the time of selection. The frequency range of interest for front–back disambiguation (11.3–16 kHz) may still have been affected by hearing loss due to many factors such as noise exposure or age,[40] and this may have affected the quantitative magnitude of our results, given that even small amounts of subclinical hearing loss have been reported to affect binaural hearing.[41–45] However, we do not find any age-related pattern in the results. We present the per-participant results for each experiment in the Appendix (Figs. 8–11), and the plots of effect-sizes vs age do not show statistically significant correlations. Further, the group-average effects are also clearly visible in the individual data from the younger participants (who also comprise most of the sample). We cannot say anything definitive about the results from the older participants (which potentially relates to hearing loss), since there were only a few of them, and our experiment was not designed to be conclusive about effects in individual participants.

Second, we instructed participants to keep their head still, but we did not monitor head movements that participants could have used to help resolve front–back confusions.

Finally, across experiments 2–4 we tested only two distractor-level regimes: just-detectable levels in experiment 2 and Zwicker loudness-matched levels in experiments 3 and 4. Our initial choice of just-detectable levels was based on pragmatic considerations, including a goal of avoiding "too loud distractors" that participants found uncomfortable, which could then lead to interactions related to attentional distraction and / or ceiling effects. When our pilot experiments indicated that distractor interference could be shown even with these just-detectable levels, we went ahead and completed the experiment with those parameters. In experiments 3 and 4, we used distractors whose loudness was equated using the Zwicker loudness function. Since the distractor stimulus with energy in the task-relevant region also falls in the less-sensitive portion of the human audiogram, the unweighted SPL of this distractor is indeed quite a bit higher than that of the other two distractors in experiments 2–4, and if the loudness relevant to this experiment is not captured by the Zwicker loudness function, it is possible that the higher unweighted SPL played a role in our results. Additional experiments with more distractor levels are required in order to see how that impacts our results presented here, in terms of the contribution of distractor energy in different spectral regions to distractor interference and in terms of the lack of evidence for informational masking from the distractor in the minimum-overlap and overlap-in-NRR conditions.

## D. Future directions

This study paves the way for further explorations of auditory motion perception in the presence of distractors. This line of research could contribute to a better understanding of multiple object tracking in hearing.

Additional experiments are required to further investigate why distractors at the front and at the right result in similar UL reductions. Other directions include manipulating the similarity between targets and distractors along other dimensions (e.g., temporal envelope, semantic content), as well as manipulating distractor salience (through variations in auditory width, spatial location, amplitude modulation, temporal evolution, etc.) and number of moving targets to investigate attentional effects.

## ACKNOWLEDGMENTS

Funded by grants from the Natural Sciences and Engineering Research Council of Canada (NSERC) Discovery Research program and Supplement (RGPIN-2022–05399 and DGECR-2022-00321 to B.S.K., RGPIN-2019–06121 and RGPAS-2019-00035 to C.G.), and the Vision Sciences Research Network (VSRN) MSc Recruitment Scholarship and a McGill Department of Physiology Richard I Birks Fellowship to N.K. We thank the Center for Interdisciplinary Research in Music Media and Technology for their facilities, as well as the technical team for their assistance. We also thank Jacky Chen for helping with proofreading the statistical results in the paper. Finally, we sincerely thank the reviewers for helping us improve the paper, and specifically for suggesting that frequency-selective attentional mechanisms could lead to informational masking playing a role in the overlap-in-RR condition.

## AUTHOR DECLARATIONS

### Conflict of Interest

The authors have no conflicts to disclose.

### Ethics Approval

The study adhered to the principles of the Declaration of Helsinki and received ethics approval from the McGill University Research Ethics Board (REB #21-03-017). Written informed consent was obtained from all participants prior to participation.

## DATA AVAILABILITY

The data that support the findings of this study are openly available in Zenodo at https://doi.org/10.5281/zenodo.22217911.

## APPENDIX

Figure 7 shows results from a preliminary pilot to adjust distractor gain for experiment 2 until the distractor was just perceptible while the target revolved at 0.9 rot/s. The top panels of Fig. 8 show per-participant results (mean and SEM for each condition across valid staircases) for experiment 1. The top panels of Fig. 9 show per-participant results (mean and SEM for each condition across four staircases) for experiment 2. Figure 10 shows per-participant data for experiment 3; there are no significant correlations between within-participant UL differences and age. Figure 11 shows per-participant data for experiment 4; there are no significant correlations between within-participant UL differences and age. Table II shows sound-pressure levels for experiment 2, rounded to the nearest 0.5 dB: target-plus-distractor in dBA (LAeq); distractor-only as dBA/dBZ pairs. Table III shows sound-pressure levels for experiment 3, rounded to the nearest 0.5 dB: target-plus-distractor in dBA (LAeq); distractor-only as dBA/dBZ pairs.

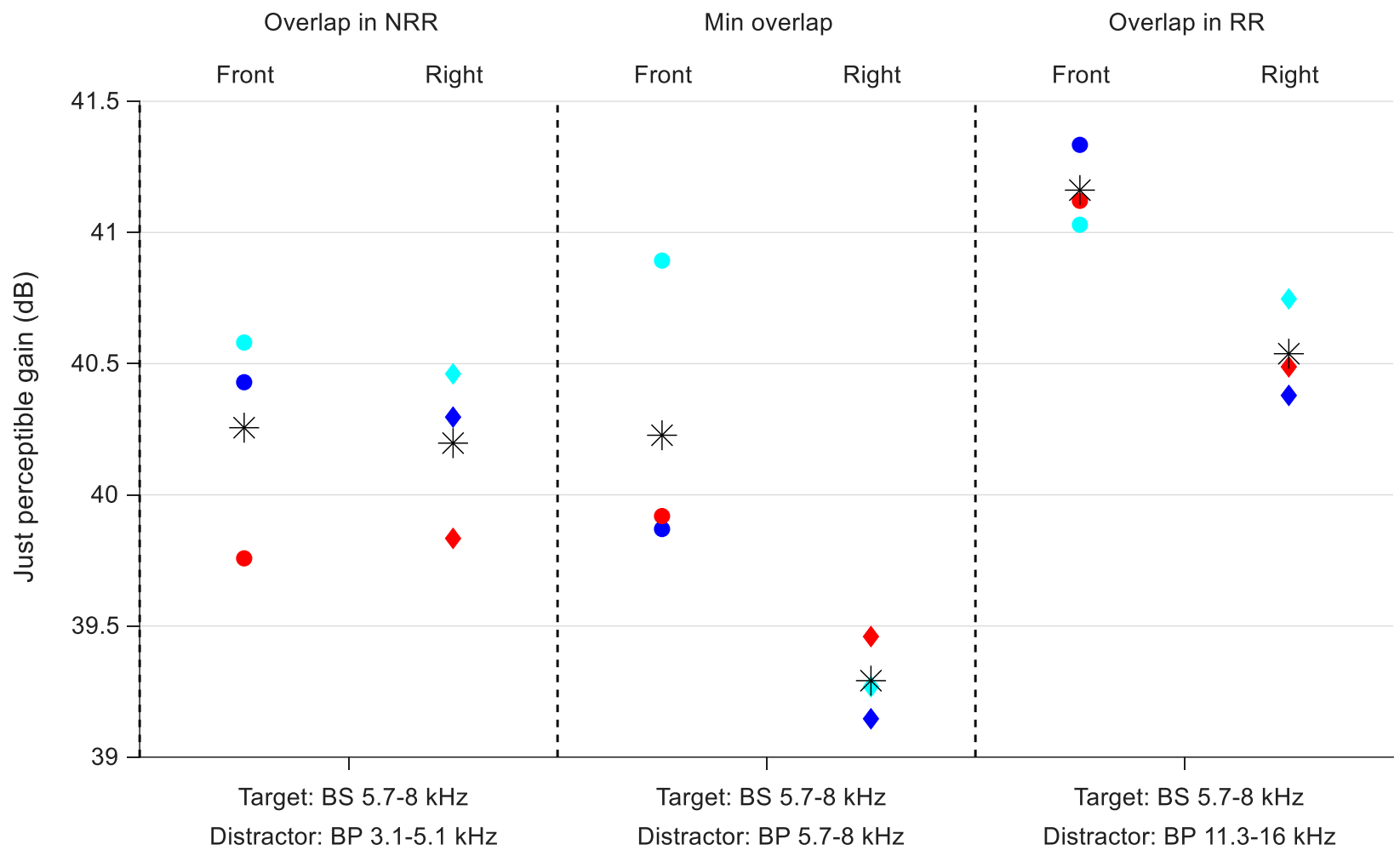

FIG. 7. Results from a preliminary pilot to adjust the gain of the distractor for experiment 2 until it is just perceptible at the resulting SPL, while the target sound was revolving around them at 0.9 rot/s. For each condition, the participant could control the gain of the distractor with a slider until it became just perceptible, and when satisfied, they confirmed, and the next trial was launched automatically. The initial signal was set at a very low level to ensure it was inaudible by default and to prevent the scaled output from reaching uncomfortably high intensities. The slider values were internally converted to decibel (dB) units relative to the digital signal amplitude. Importantly, these values reflect relative amplitude scaling within the experiment rather than absolute sound pressure levels measured in the room. Data points in three colors show data from three experienced participants, and each participant completed eight trials (four CW; four CCW). Circles are for a distractor at the front and diamonds for a distractor on the right. The black asterisk is the mean across the three participants for each condition, and the condition in each column is identified by the text on top.

TABLE II. SPL of the stimuli presented in experiment 2, rounded to the nearest 0.5 dB. Target-plus-distractor levels are reported in dBA (LAeq). Distractor-only levels are reported as dBA/dBZ pairs, with the A-weighted value first and the Z-weighted value second. After the "just perceptible" pilot loudness adjustment, the selected scaling values in Fig. 7 were applied in the main experiment, and the resulting sound levels that were measured in the room are shown here.

| Condition | SPL target + distractor (dBA) | SPL distractor (dBA/dBZ) |
|---|---|---|
| Control | 42 | - |
| Overlap in NRR | 42.5 | 30.5/29.5 |
| Minimum overlap | 42.5 | 29/29.5 |
| Overlap in RR | 42.5 | 28.5/33.5 |

TABLE III. Stimuli presented during experiment 3 and their sound pressure level, rounded to the nearest 0.5 dB. Target-plus-distractor levels are reported in dBA (LAeq). Distractor-only levels are reported as dBA/dBZ pairs, with the A-weighted value first and the Z-weighted value second. Levels were measured after a loudness adjustment with the Zwicker loudness model. Same format as in Table II.

| Condition | SPL target + distractor (dBA) | SPL distractor (dBA/dBZ) |
|---|---|---|
| Control | 42 | - |
| Overlap in NRR | 42.5 | 30.5/29.5 |
| Minimum overlap | 42.5 | 32.5/33 |
| Overlap in RR | 43.5 | 39/44 |

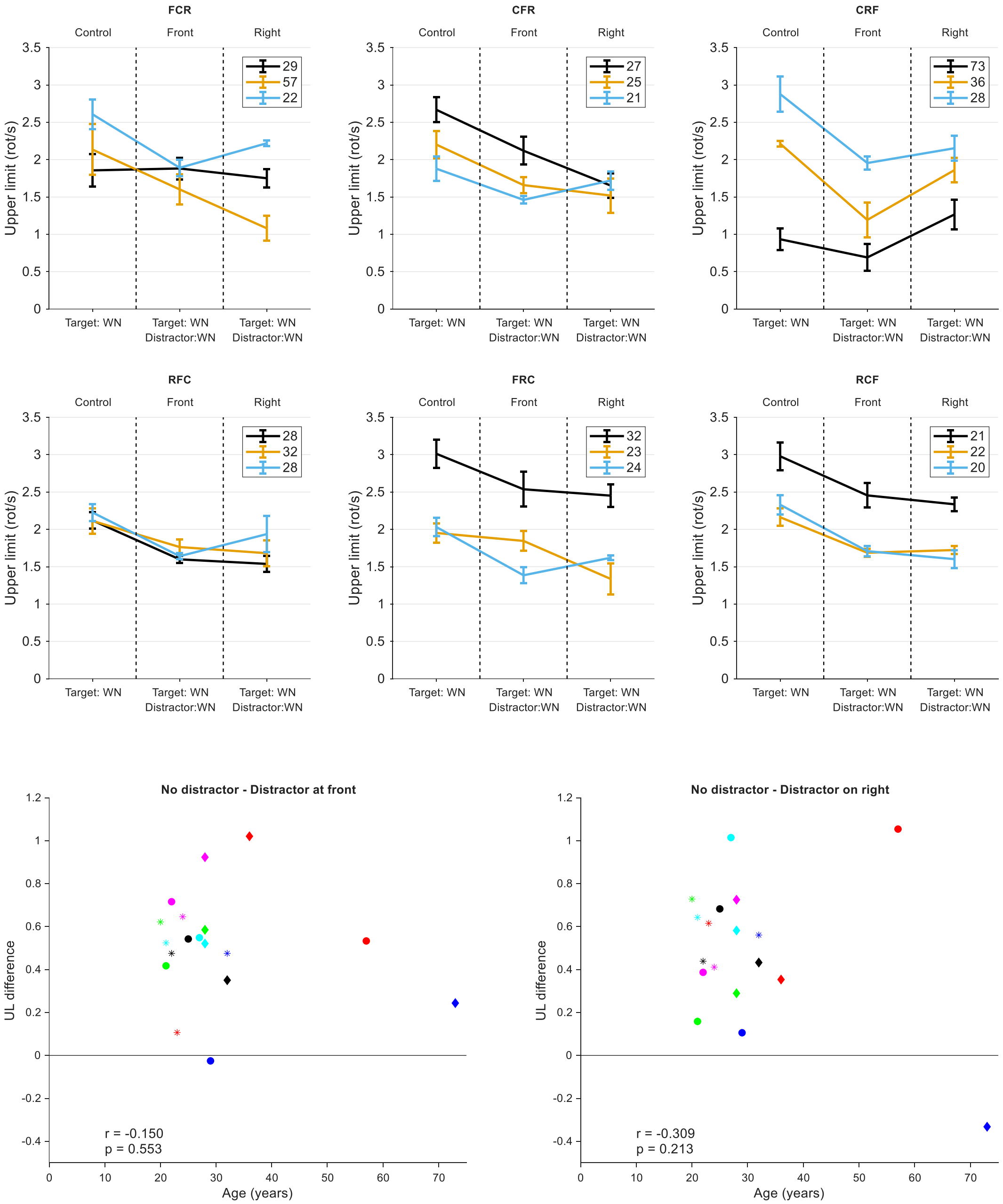

FIG. 8. Top panels show per-participant results for experiment 1: mean and SEM for each condition across valid staircases (normally four, except for p17, where the control and right conditions contained three and two valid staircases, respectively). Each panel shows the results from the three participants who performed the three blocks in a particular order: C, control; F, distractor at front; R, distractor at right. The legend shows participant ages. Bottom panels show the difference between ULs for control and distractor conditions (at front or at right). The same symbol-color combination on the two plots refers to the same subject. No correlation of UL difference with age was found (Pearson's $r$ and $p$-value shown in plots). The differences mostly lie above zero, reflecting the robustly higher ULs in the control condition compared to the distractor conditions.

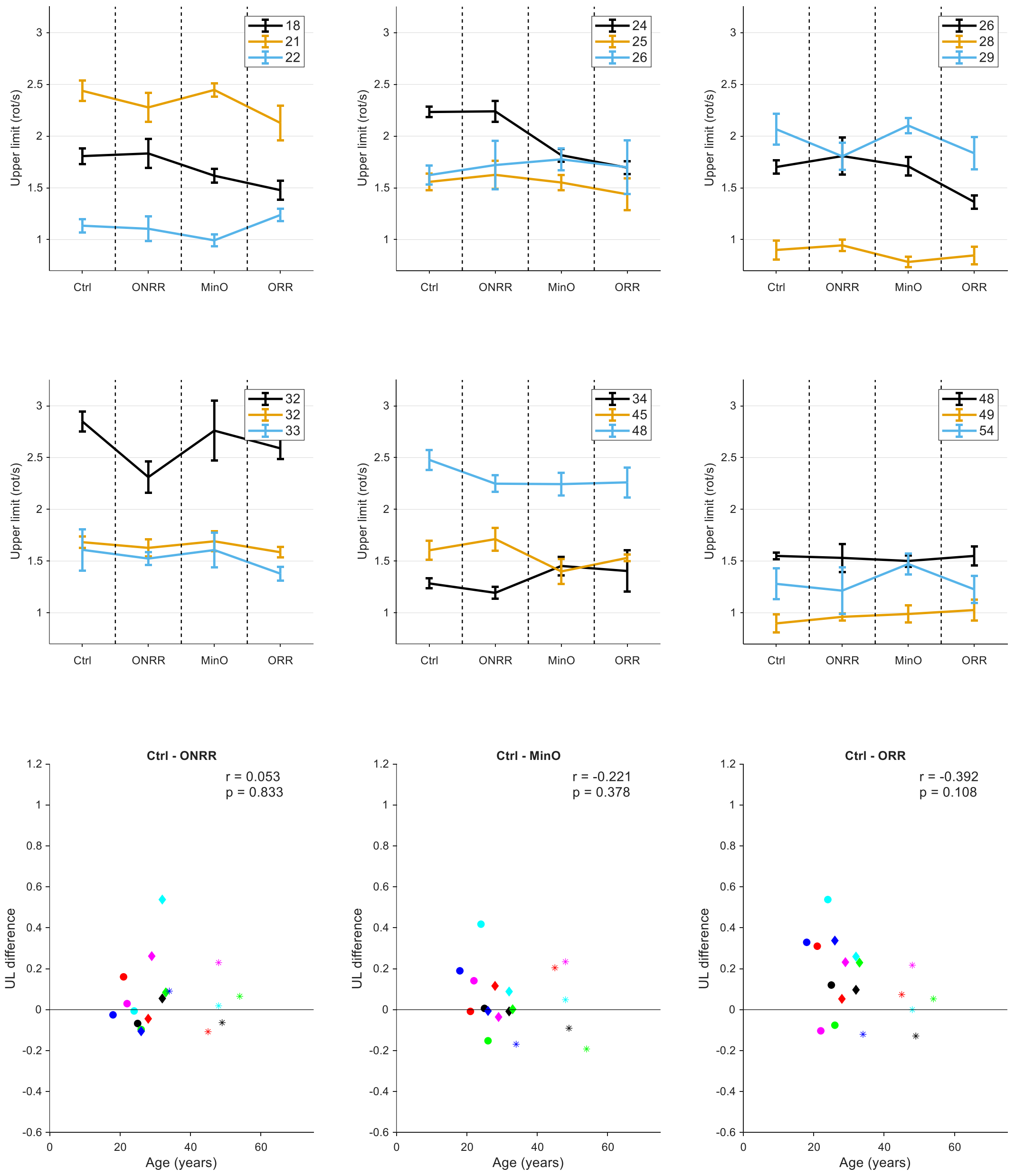

FIG. 9. Top panels show per-participant results (mean and SEM for each condition across the four staircases) for experiment 2. Each panel shows the results from three participants, sorted by age from left to right, top to bottom. The legend shows participant ages. Each participant performed a different block order for the four conditions. Bottom panels show the difference between ULs for control and distractor conditions (overlap-NRR, min-overlap and overlap-RR). No correlation of UL difference with age was found (Pearson's *r* and *p*-value shown in graph).

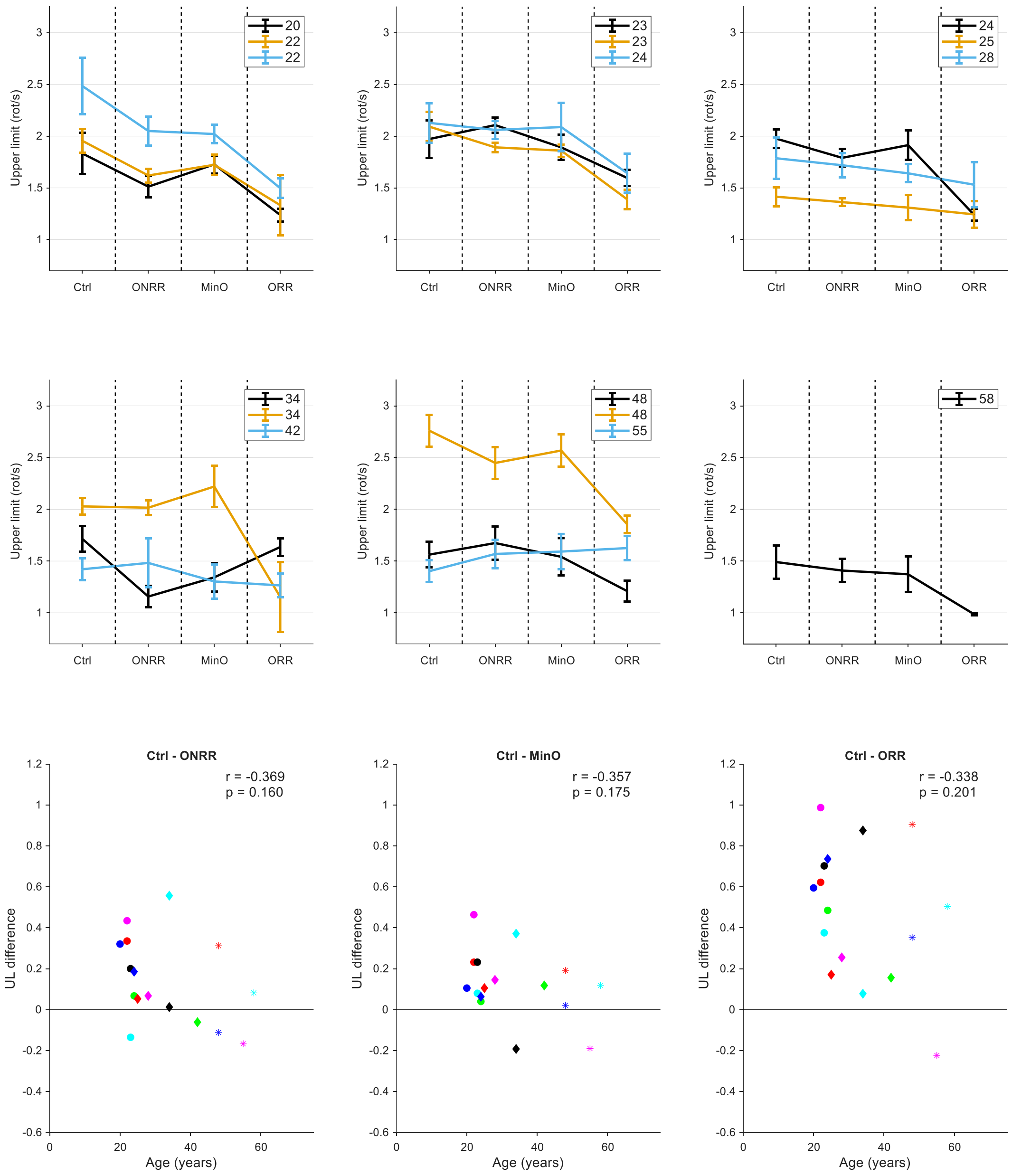

FIG. 10. Per-participant data for experiment 3, and a lack of correlation of UL difference with age. Same format as in Fig. 9. One participant did not provide consent to make individual data available.

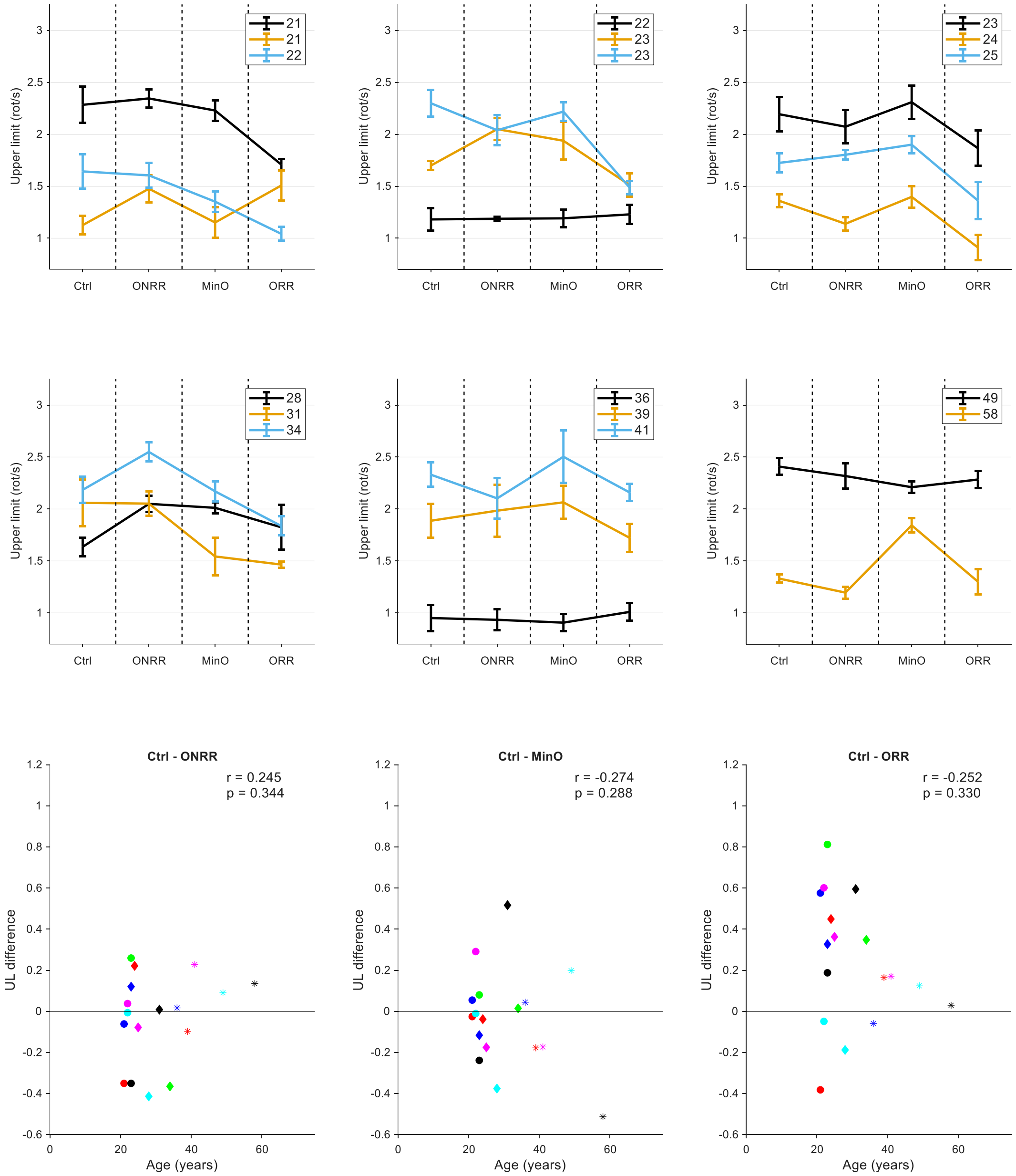

FIG. 11. Per-participant data for experiment 4, and a lack of correlation of UL difference with age. Same format as in Fig. 9.